\documentclass[a4paper]{jpconf}

\usepackage{graphicx}

\usepackage{hyperref}

\usepackage[sort&compress,numbers]{natbib}

\usepackage{amsmath}
\usepackage{amssymb}
\usepackage{bm}

\newcommand{\maestro}{{\sffamily Maestro}}
\newcommand{\maestroex}{{\sffamily MAESTROeX}}
\newcommand{\castro}{{\sffamily Castro}}
\newcommand{\nyx}{{\sffamily Nyx}}
\newcommand{\amrex}{{\sffamily AMReX}}
\newcommand{\starkiller}{{\sffamily Starkiller Microphysics}}
\newcommand{\amrexastro}{{\sffamily AMReX-Astro}}
\newcommand{\fboxlib}{{\sffamily FBoxLib}}
\newcommand{\boxlib}{{\sffamily BoxLib}}

\newcommand{\Ub}{{\,\bm{U}}}

\newcommand{\pd}[2]{\frac{\partial #1}{\partial #2}}
\newcommand{\md}[2]{\frac{\mathrm{D} #1}{\mathrm{D} #2}}

\usepackage{color}
\begin{document}

\title{Modelling low Mach number stellar hydrodynamics with MAESTROeX}

\author{A.~Harpole$^1$,
        D. Fan$^2$,
        M.~P. Katz$^3$,
        A.~J. Nonaka$^2$,
        D.~E. Willcox$^2$, and
        M. Zingale$^1$}

\address{$^1$Department of Physics and Astronomy, Stony Brook
  University, Stony Brook, NY 11794-3800 USA}

\address{$^2$Center for Computational Sciences and Engineering,
  Lawrence Berkeley National Lab, Berkeley, CA 94720 USA}

\address{$^3$NVIDIA Corporation, 2788 San Tomas Expressway,
  Santa Clara, CA, 95050 USA}

\ead{alice.harpole@stonybrook.edu}


\begin{abstract}
Modelling long-time convective flows in the interiors of stars is extremely challenging using conventional compressible hydrodynamics codes due to the acoustic timestep limitation.
Many of these flows are in the low Mach number regime, which allows us to exploit the relationship between acoustic and advective time scales to develop a more computationally efficient approach.
\maestroex\ is an open source low Mach number stellar hydrodynamics code that allows much larger timesteps to be taken, therefore enabling systems to be modelled for much longer periods of time. This is particularly important for the problem of convection in the cores of rotating massive stars prior to core collapse. To fully capture the dynamics, it is necessary to model these systems in three dimensions at high resolution over many rotational periods.  We present an overview of \maestroex's current capabilities, describe ongoing work to incorporate the effects of rotation and discuss how we are optimising the code to run on GPUs. 
\end{abstract}


\section{Introduction} \label{sec:intro}

For many flows in astrophysical systems, the magnitude of the fluid velocity is much less than the soundspeed. Consequently, the Mach number, the ratio of the fluid velocity to the sound speed, is much less than 1: $\textrm{Ma} = |\Ub| / c_s \ll 1$. Such low Mach number flows are challenging to model with standard compressible schemes using explicit timestepping methods, where the maximum size of the timestep is determined by the Courant-Friedrich-Lewy (CFL) condition. This states that for a grid with cell spacing $\Delta x$, the timestep $\Delta t <  \Delta x / \max(|\Ub|+c_s)$. For low Mach number flows, this condition is dominated by the contributions of the sound speed, with the result that many fine timesteps are required when using high spatial resolution. Long timescale, high resolution simulations therefore become extremely computationally expensive. 

This restriction can be lifted by using sound-proof methods. These methods can involve modifying the fluid equations, modifying the computational algorithm and/or modifying the flow parameters in order to allow much larger timesteps to be used. In \maestroex, we use the \emph{low Mach number approximation}, a limit of the compressible Euler equations which effectively filters out sound waves. An overview of this shall be given in \sref{sec:low_mach_hydro}, with further details of the \maestroex~code and our development workflow given in \sref{sec:maestroex} and \sref{sec:workflow}.  

An example of an astrophysical low Mach number flow is convection within the interiors of massive stars. Prior to core collapse, the interiors of massive stars consist of a sequence of convective burning shells, separated by inert, non-convective shells. Moving from the outermost layers inwards, these shells consist of heavier and heavier elements, with the core burning elements up to ${}^{56}$Fe. An accurate description of the structure of such stars in the minutes before core collapse is important for supernova modelling. The composition and structure of the star prior to collapse provides the initial data for these models, so this needs to be accurate if the resulting supernova models are to be trusted. Modelling this convection is challenging using conventional compressible schemes because the domain size is very large (i.e.~a significant fraction of the interior of the entire star), high resolution is required in order to properly capture the turbulent mixing that occurs at shell boundaries, and long time periods are required (multiple convective turnover times). In \sref{sec:rotation}, we shall describe ongoing work to model this problem in \maestroex, in particular describing how we are adding rotation to our existing scheme. 

The latest generation of supercomputers coming online are relying more and more on GPU architectures in order to increase performance whilst minimizing power requirements. For HPC codes to best exploit the most powerful machines, it is therefore becoming increasingly necessary to port these codes so that they can run on GPUs. In \sref{sec:gpus}, we describe our work to port \maestroex~to run on GPUs and demonstrate its performance.


\section{Low Mach number hydrodynamics} \label{sec:low_mach_hydro}

Sound-proof methods for modelling low Mach number flows can take a variety of forms. One technique is to use preconditioners in order to reduce the stiffness of the equations, allowing larger timesteps to be used \cite{Miczek2014,Barsukow2016}. Another technique is to modify the flow parameters, artificially boosting the speed of the flow without changing its behaviour so that the system evolves faster and fewer timesteps are required \citep{Rempel2005,Hotta2012}. Fully implicit time integration codes, such as those used by \cite{Viallet2011,Viallet2015,Goffrey2016}, are no longer restricted by the CFL condition and so can use arbitrarily large timesteps.

The technique that we use in \maestroex~is to modify the fluid equations themselves so as to filter out the soundwaves. This is a similar approach to the incompressible \cite{Boussinesq1901} and anelastic \cite{Ogura1962a,Gough1968,Durran1989} approximations, however the approximation that we use (called the low Mach number approximation \cite{Day2000,Almgren2006a,Nonaka2010} and the generalised pseudo-incompressible approximation \cite{Vasil2013}) allows for background stratification and large density and temperature perturbations due to heating and changes in composition. This is achieved by decomposing the pressure into a one-dimensional hydrostatic base state, $p_0 = p_0(r, t)$, and a dynamic pressure perturbation, $\pi = \pi(\bm{x}, t)$, such that the full pressure is given by $p(\bm{x}, t) = p_0(r, t) + \pi(\bm{x}, t)$. In the low Mach number regime, asymptotic analysis shows that $|\pi|/p_0 = O(\mathrm{Ma}^2)$. 

The low Mach number fluid equations are given by 
\begin{align}
    \pd{\left(\rho X_k\right)}{t} &= - \nabla\cdot\left(\rho X_k \Ub \right) + \rho\dot{\omega}_k, \\
    \pd{\Ub}{t} &= - \Ub\cdot\nabla\Ub - \frac{\beta_0}{\rho}\nabla\left(\frac{\pi}{\beta_0}\right) - \frac{\rho - \rho_0}{\rho} g \bm{e}_r,\\
    \pd{\left(\rho h\right)}{t} &= -\nabla\cdot\left(\rho h \Ub\right) + \md{p_0}{t} + \rho H_{\mathrm{nuc}},
\end{align}
where $\rho$, $\Ub$ and $h$ are the mass density, fluid velocity and specific enthalpy, $X_k$ and $\dot{\omega}_k$ are the species mass fraction and production rate of species $k$, and $H_{\mathrm{nuc}}$ is the energy release per time per unit mass. \maestroex~defines the base state pressure $p_0$ to be consistent with the one-dimensional base state density, $\rho_0 = \rho_0(r, t)$, which represents the lateral average and is in hydrostatic equilibrium with $p_0$:
\begin{equation}
    \nabla p_0 = -\rho_0 g \bm{e}_r,
\end{equation}
where $g = g(r,t)$ is the magnitude of the gravitational acceleration, and $\bm{e}_r$ is the radial unit vector. The background stratification is captured by introducing a buoyancy-like term, $\beta_0$. It is defined as 
\begin{equation}
    \beta_0(r,t) = \rho_0(0,t) \exp\left(\int_0^r dr'\, \frac{1}{\overline{\Gamma}_1p_0}\pd{p_0}{r'} \right),
\end{equation}
where $\overline{\Gamma}_1$ is the lateral average of the first adiabatic exponent, $\Gamma_1 \equiv d(\ln p)/d(\ln \rho)|_s$, and $s$ is the entropy. The equations are closed by casting the equation of state (EoS) as a velocity divergence constraint. This is done by taking the Lagrangian derivative of the EoS for pressure as a function of the thermodynamic variables, substituting in the equations of motion for mass and energy, and requiring that the pressure is described by a function of $r$ and $t$ based on the condition of hydrostatic equilibrium. Details of this derivation can be found in \cite{Almgren2006a,Almgren2006b}. This constraint is given by 
\begin{equation}
    \nabla\cdot\left(\beta_0\Ub\right) = \beta_0 \left(S - \frac{1}{\overline{\Gamma}_1 p_0}\pd{p_0}{t} \right).
\end{equation}
Here, $S$ is an expansion term which describes local compressibility effects due to changes in composition and heating from reactions:
\begin{equation}
    S = -\sigma \sum_k \xi_k\dot{\omega}_k + \frac{1}{\rho p_\rho}\sum_k p_{X_k}\dot{\omega}_k + \sigma H_{\mathrm{nuc}},
\end{equation}
where we define the following thermodynamic quantities as
\begin{align*}
    p_{X_k} \equiv \left.\pd{p}{X_k}\right|_{\rho,T,X_{j,j\neq k}},\qquad
     \xi_k&\equiv \left.\pd{h}{X_k}\right|_{p, T,X_{j,j\neq k}},\qquad
     p_\rho\equiv \left.\pd{p}{\rho}\right|_{T, X_k},\\
     \sigma \equiv \frac{p_T}{\rho c_p p_\rho}, \qquad
     p_T&\equiv \left.\pd{p}{T}\right|_{\rho, X_k} \quad\mathrm{and}\quad 
     c_p\equiv \left.\pd{h}{T}\right|_{p, X_k}.
\end{align*}


\section{MAESTROeX} \label{sec:maestroex}

Prior to \maestroex, we developed the low Mach number code \maestro. Like \maestroex, \maestro~is a block-structured adaptive mesh refinement (AMR) code for modelling low Mach number astrophysical codes. Its development is described in a series of papers: \cite{Almgren2006a,Almgren2006b,Almgren2008a,Zingale2009,Nonaka2010}. In \maestro, the system of low Mach number equations is solved using an explicit Godunov approach for the advection, a stiff ODE solver for the reactions (VODE~\cite{vode}), and multigrid linear solvers for the pressure projection steps. Strang splitting~\cite{strang:1968} is used to integrate the thermodynamic variables, a second order projection method to integrate the velocity subject to the divergence constraint, and a velocity splitting scheme to hydrodynamically evolve the base state. The original \maestro~code was developed using the Fortran 90 interface of the \boxlib~software framework \cite{Zhang2016}; \maestroex~instead uses the C++/Fortran 90 \amrex~framework \cite{Zhang2019}. 

\maestro~has been used to model a number of astrophysical systems, including convection in white dwarfs prior to type Ia supernovae \cite{Zingale2011,Nonaka2011,Malone2014a,Zingale2013,Jacobs2016}, convection in massive stars \cite{Gilet2013} and type I X-ray bursts \cite{Malone2011,Malone2014,Zingale2015}. 

The numerical algorithm implemented in \maestroex~improves upon the original \maestro~algorithm in a number of ways. The temporal integration method has been greatly simplified without compromising the second order accuracy, and a new spherical base state mapping has been implemented in order to reduce mapping errors between spherical and Cartesian grids. Utilising the \amrex~framework has allowed us to implement MPI+OpenMP (with tiling \cite{Zhang2016}) parallelism, which has been shown to scale well to over 10,000 MPI processes. 

Further details of the algorithm implemented in \maestroex~and its performance can be found in \cite{Fan2019}.


\section{AMReX-Astro development workflow} \label{sec:workflow}

\maestroex~is part of the \amrexastro~suite of open source adaptive mesh refinement hydrodynamics codes for astrophysical flows. Other codes in this family include \castro, an astrophysical radiation hydrodynamics simulation code \cite{Almgren2010}, and \nyx, an N-body hydrodynamics cosmological simulation code \cite{Almgren2013}. All three codes are developed using the \amrex~software framework, and share much in common in terms of their development, structure and numerical methods. \castro~and \maestroex~share a common set of microphysics solvers provided by the \starkiller~library\footnote{\url{https://github.com/starkiller-astro/Microphysics}}. Developers of the different codes work closely together (in fact many of the developers work on more than one of the codes), with their shared structure and framework meaning that new features developed in one code can easily be replicated in the others. A prime example of this is the GPU porting capability that has recently been implemented in \castro, and is currently in the process of being implemented in \maestroex. 

All codes in the \amrexastro~suite are open source, with all development carried out in public repositories hosted on GitHub\footnote{\url{https://github.com/AMReX-Astro}}. We believe that having both the codes and the development process completely open promotes good scientific practices, as it means that results from our simulations are reproducible and the codes used to produce them can be examined and improved by the community.  Others can use the codes for their own scientific investigations, and it is possible to adapt all or parts of the codes to suit new problems. Using version control allows us to keep a record of the codes' development process, helping us to track down bugs and means that new and existing developers can learn from previous mistakes. The development branches of the codes are tested nightly using a test suite of problems, checking that any new additions to the code have not significantly changed any of the solutions or slowed down the performance.

New versions of the codes are released on the first of each month. In the case of \castro, these versions are then archived on Zenodo\footnote{\url{https://zenodo.org/}}. This also provides us with DOIs (digital object identifiers), further enhancing the reproducibility of our results and ensuring the sustainability of the codes used to produce them. We intend to extend this archiving procedure to the other \amrexastro~codes in the near future.


\section{Rotation} \label{sec:rotation}

\maestroex~is currently able to model spherically symmetric systems, however it has no support for rotation which breaks the spherical symmetry. We wish to model convection in the interiors of massive stars, and it is known that these stars often have non-negligible rotational frequencies and that this rotation can have have significant effect on mixing at convective shell boundaries. We are therefore currently exploring several possible ways of implementing rotation in \maestroex. 

As described in \cite{Zingale2011}, to incorporate rotation in our equation set, we add the Coriolis and/or centrifugal terms to the velocity evolution equation:
\begin{equation}
    \pd{\Ub}{t} = -\Ub\cdot\nabla\Ub - \frac{\beta_0}{\rho}\nabla\left(\frac{\pi}{\beta_0}\right) - \frac{\rho - \rho_0}{\rho} g \bm{e}_r - 2\boldsymbol{\Omega} \times \Ub - \boldsymbol{\Omega} \times (\boldsymbol{\Omega} \times \bm{r}),
\end{equation}
where $\boldsymbol{\Omega}$ is the angular velocity, $\bm{F}_\text{Coriolis}=- 2\boldsymbol{\Omega} \times \Ub$ is the Coriolis force and $\bm{F}_\text{centrifugal}= - \boldsymbol{\Omega} \times (\boldsymbol{\Omega} \times \bm{r})$ the centrifugal force.  Note that for slowly rotating systems it is a reasonable approximation to ignore the centrifugal force.

Another way that rotation could be incorporated would be to introduce a new `rotational pressure term', $p_1$, in order to balance the centrifugal potential:
\begin{equation}
    \nabla p_0 + \nabla p_1 = \rho \nabla \phi + \frac{1}{2}\rho \nabla\Omega^2 r^2 = \rho \nabla \Phi_{\text{eff}},
\end{equation}
where $\phi = \phi(r)$ is the gravitational potential (so $\nabla \phi(r) = -g \bm{e}_r$), $\Omega$ is the rotational frequency, and $\Phi_{\text{eff}}$ is the new effective gravitational potential. 

In \maestroex, we currently write the one-dimensional base state as a function of the radial coordinate, $r$. For a spherical system, this therefore assumes the base state to be spherically symmetric. However, a star that is rotating at a significant rate will no longer be spherical: it will instead become oblique, bulging around the equator. In order to capture the star's deformation without sacrificing the base state, we can rewrite the base state pressure as a function of the effective potential: $p_0(r)\rightarrow p_0(\Phi_{\text{eff}})$. We can then rewrite the other base state quantities and the system of low Mach number equations as functions of the new coordinates $(\Phi_\text{eff}, \bm{x}, t)$.

The massive stars that we are interested in modelling are believed to be relatively slow rotators, with the rotation rate in the core prior to collapse no more than a few percent of the Kepler frequency \cite{heger2004presupernova}, so it is likely that one of the simpler approaches outlined above should be sufficient. However, for modelling e.g.~type I X-ray bursts, it may be that a more sophisticated approach is needed. Type I X-ray bursts are produced by thermonuclear deflagrations in the liquid surface layers of fast rotating neutron stars. These stars typically rotate at frequencies of $\sim$300\textendash600 Hz, a significant fraction of the stars' breakup velocity. This produces non-negligible deformation of the star about the equator, and may cause surface flows with a significant dependence on the meridional angle. To properly capture the effects of rotation in such a system, it may therefore be necessary to use a two-dimensional base state, with the base state pressure a function of both the radius and the meridional angle $p_0(r)\rightarrow p_0(r, \theta)$.


\section{GPUs} \label{sec:gpus}

The latest generation of supercomputers to come online (e.g.\ OLCF Summit, soon NERSC Perlmutter and OLCF Frontier) rely heavily on GPUs in order to achieve high performance while keeping energy consumption to a minimum (GPUs consume considerably less power per flop than CPUs). This trend is set to continue: supercomputers are moving to new architectures in order to increase performance rather than simply adding more and more CPUs. In order for HPC codes to be able to exploit these latest machines, it is therefore becoming increasingly necessary to port codes to run on GPUs.

We have begun porting \maestroex~to GPUs, leveraging the GPU capabilities built into \amrex~and following the lessons learned porting \castro~to GPUs. In the \amrex~framework, routines can be ported to GPUs by simply inserting a few macros into the code. The boilerplate code required to compile this for GPU is then generated by a custom preprocessor prior to compilation. CUDA managed memory is used to take care of data transfers between the CPU and GPU, so no explicit copies of data to/from the CPU and GPU are required, significantly reducing the extra code required to port the code. One of the key emphases of the \amrex~approach to GPUs is to maintain performance portability: porting code to GPUs should not come at the expense of the code's performance on CPU-only machines. Because GPU-specific code is created by the preprocessor prior to compilation, it is not necessary to maintain separate versions of the code for different architectures. By changing the compiler flags, the same bit of code can be compiled for serial, MPI-only, OpenMP-only, MPI+OpenMP, GPU-only or MPI+GPU.
In order to achieve the best performance on the GPU, we require some additional optimisation of the computational kernels. However, we find that these optimisations often improve the performance when the code is run on the CPU only as well. 
More details of the approach used to port our codes to GPU are to appear in a forthcoming \amrexastro~GPU paper. 

So far we have ported most of the source terms and hydrodynamics in \maestroex~to run on GPUs. GPU support for the linear solvers is becoming increasingly available from the \amrex~team.  We have found that one of the challenges moving forwards is dealing with the one-dimensional base state. Currently, each MPI process holds a copy of the entire base state. As the problem size and/or resolution gets larger, the size of the base state grows. Data transfers between the CPU and GPU are typically a slow process, so as the base state grows in size, the cost of these transfers becomes more significant and reduces the performance overall. As we are still in the process of porting our code to GPUs, there still remain some functions operating on the base state which run on the CPU. Every time one of these is called, if the necessary base state data is currently on the GPU, it is copied back to the CPU (and then back again next time it is required for a GPU-based calculation).

To alleviate this problem, we are looking at ways to reduce both the number and size of these data transfers. For example, in some routines we map the base state to the three-dimensional grid \textendash~for these, we can move this operation to the CPU so that the GPU does not require the entire 1d base state to be copied over. It may be necessary for us to develop a way to split up the base state, so that subgrids only receive the sections of the base state necessary for their calculations (rather than the base state for the entire problem domain). 

\begin{figure}
    \centering
    \includegraphics[width=\textwidth]{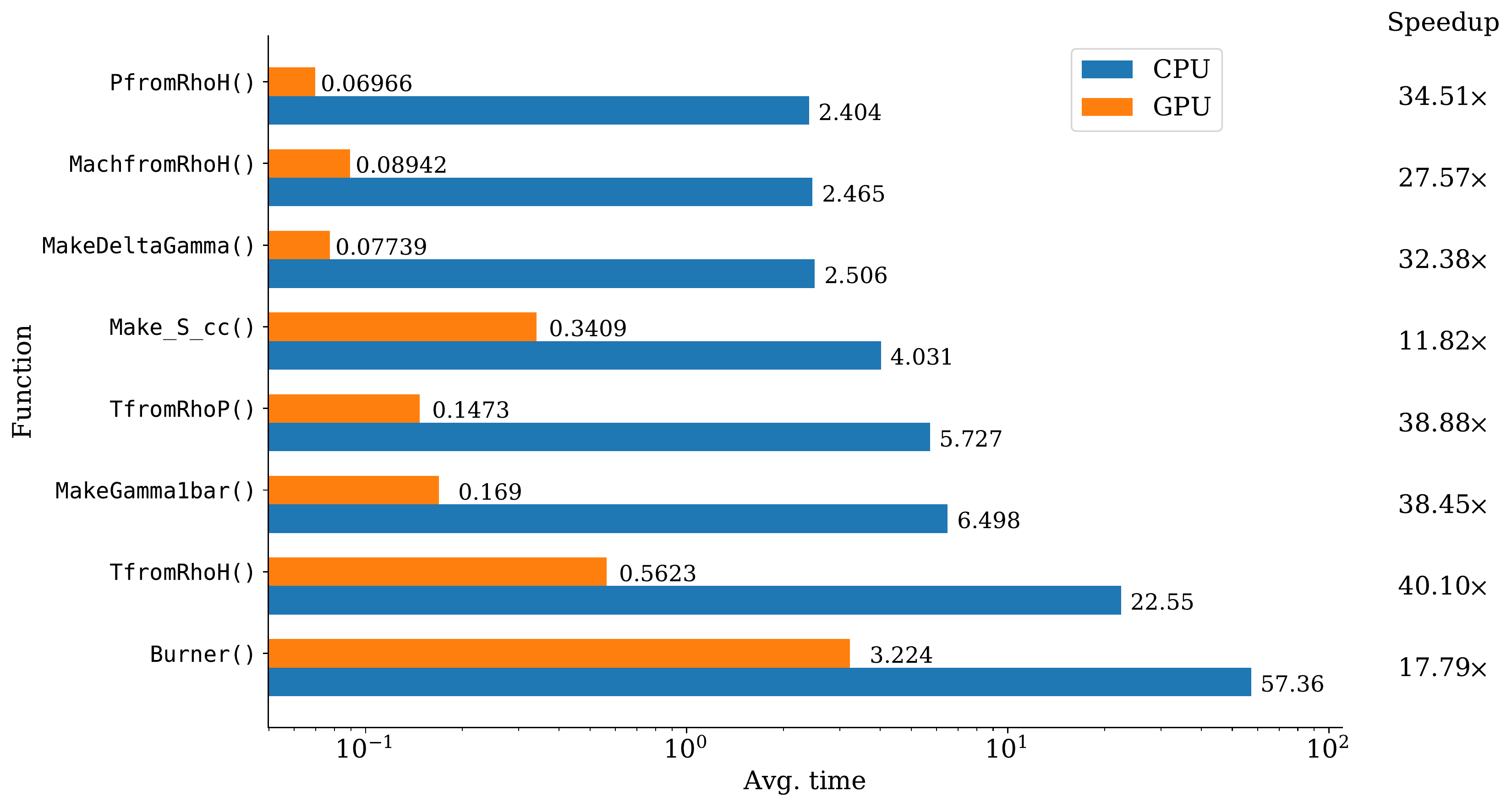}
    \caption{Comparison of the performance of functions on the CPU vs the GPU. The plot shows the average execution time for several functions, recorded for a run of the three-dimensional \texttt{reacting\_bubble} problem on Summit. Both tests were performed using a single node, with the CPU test using 42 MPI processes, each with 4 OpenMP threads, and the GPU test using 6 MPI processes, each process associated with a single GPU. Note that the $x$-axis scale is logarithmic. On the right, we have listed the speedup for each function when run on GPUs.}\label{fig:gpu_speedup}
\end{figure}

As mentioned, our work porting \maestroex~to run on GPUs is still ongoing, however in Figure~\ref{fig:gpu_speedup} we show the speedup we have achieved so far for a number of individual functions that have been offloaded to GPU. The plot shows the average execution time of each function, recorded for the three-dimensional \texttt{reacting\_bubble} problem on the OLCF Summit machine.
This problem uses a simple reaction network modelling $^{12}\mathrm{C} + {}^{12}\mathrm{C} \rightarrow {}^{24}\mathrm{Mg}$ that can be integrated on the GPUs using the CUDA Fortran port of VODE \citep{vode} described in \cite{Zingale2018}.
Both tests were run on a single node, consisting of two IBM POWER9 processors (together providing 42 physical CPU cores) and six NVIDIA Volta GPUs, and compiled using the PGI compiler version 19.4. The CPU test used 42 MPI processes, each with 4 OpenMP threads, and the GPU test used 6 MPI processes, each process associated with a single GPU.
Note that the plot's $x$-axis scale is logarithmic. As can be seen, large speedups in excess of 30-40$\times$ were achieved for several of the functions when run on the GPU. Once we have finished porting the remaining routines to GPU and further optimised our code, we hope to achieve this sort of speedup for the entire timestep.

\section{Summary} \label{sec:summary}

\maestroex~is an open source code for modelling low Mach number astrophysical flows. It uses the low Mach number approximation, which allows it to use a much larger timestep than conventional compressible schemes by filtering out the soundwaves. Unlike other sound-proof methods, we still retain background stratification and large density and temperature perturbations due to local compositional changes and heating, both of which are particularly important for modelling atmospheres and burning. 

\maestroex~is a new and improved version of our previous code \maestro. It is based on \amrex~(rather than \fboxlib), which allows us to exploit its powerful new solvers and make use of ongoing improvements. \maestroex~is part of the \amrexastro~suite of open source adaptive mesh refinement astrophysical hydrodynamics codes (which also includes \castro~and \nyx). New features developed in other codes in the family can therefore be easily implemented in \maestroex. As development is open, results from our simulations can be reproduced, and monthly versioning ensures software sustainability.

We're currently working on implementing rotation in \maestroex. For modelling the interiors of massive stars (which rotate relatively slowly), we may be able to get away with neglecting the centrifugal force and simply add the Coriolis force as a source term. However, this will not be sufficient for faster rotating systems (e.g.~for modelling Type I X-ray bursts on the surfaces of millisecond pulsars), so we are also exploring other methods including modelling the centrifugal force as an effective pressure term, rewriting the base state as a function of the effective gravitational potential and using a two-dimensional base state which is also a function of the meridional angle.

We're currently in the process of porting \maestroex~to run on GPUs, using our experience doing the same for \castro. This will enable us exploit the latest supercomputer architectures. So far, we've ported source terms and are in the process of porting the hydro. 

Work has recently begun on implementing the new time integration strategy, spectral deferred corrections (SDC) that has recently been implemented in \castro~\cite{castro:sdc}. This method eliminates the coupling error between source terms and hydrodynamics incurred in operator splitting techniques. It is hoped that this shall be particularly useful for our simulations of rotating massive stars, where we have found energetic reactions in the star's core to be challenging to model using our current methods.


\ack Figure~\ref{fig:gpu_speedup} was generated using \texttt{Jupyter}
\citep{Kluyver2016} and \texttt{matplotlib} \citep{Hunter2007}.  The
work at Stony Brook was supported by the SciDAC program DOE grant
DE-SC0017955 and DOE/Office of Nuclear Physics grant
DE-FG02-87ER40317.  The work at LBNL was supported by the DOE Office
of Advanced Scientific Computing Research under Contract No,
DE-AC02-05CH11231.  An award of computer time was provided by the
Innovative and Novel Computational Impact on Theory and Experiment
(INCITE) program. This research used resources of the Oak Ridge
Leadership Computing Facility at the Oak Ridge National Laboratory,
which is supported by the Office of Science of the U.S. Department of
Energy under Contract No.\ DE-AC05-00OR22725.  This research used
resources of the National Energy Research Scientific Computing Center,
which is supported by the Office of Science of the U.S. Department of
Energy under Contract No.\ DE-AC02-05CH11231. This research has made
use of NASA's Astrophysics Data System Bibliographic Services.

\bibliographystyle{iopart-num}
\bibliography{ws}

\end{document}